%% file: sample-sigconf.tex
\documentclass[sigconf,natbib=true,anonymous=false]{acmart}

\usepackage{color}
\usepackage{graphicx}
\usepackage{float}
\usepackage{subfigure}
\usepackage{threeparttable}
\usepackage{bm}

\usepackage{pifont}
\usepackage{booktabs}

\usepackage{ulem}

\usepackage{multirow}

\usepackage{enumitem}
\usepackage{ulem}
\usepackage{rotating}

\theoremstyle{definition}

\usepackage{amsmath}

\usepackage{graphicx}
\usepackage{extarrows}
\usepackage{enumitem}

\AtBeginDocument{%
  \providecommand\BibTeX{{%
    \normalfont B\kern-0.5em{\scshape i\kern-0.25em b}\kern-0.8em\TeX}}}

\copyrightyear{2024}
\acmYear{2024}
\setcopyright{acmlicensed}
\acmConference[SIGIR '24] {Proceedings of the 47th International ACM SIGIR Conference on Research and Development in Information Retrieval}{July 14-18, 2024}{Washington D.C., USA}
\acmBooktitle{Proceedings of the 47th International ACM SIGIR Conference on Research and Development in Information Retrieval (SIGIR '24), July 14-18, 2024, Washington D.C., USA}
\acmPrice{15.00}
\acmISBN{978-1-4503-9408-6/23/07}
\acmDOI{10.1145/XXXXXX.XXXXXX}

\begin{document}

\title{CaDRec: Contextualized and Debiased Recommender Model}
\author{Xinfeng Wang}
\email{g22dtsa7@yamanashi.ac.jp}
\affiliation{%
\institution{University of Yamanashi}
\orcid{0000-0003-4491-8369}
\city{Kofu}
\country{Japan}
}

\author{Fumiyo Fukumoto}
\email{fukumoto@yamanashi.ac.jp}
\affiliation{%
  \institution{University of Yamanashi}
  \orcid{0000-0001-7858-6206}
\city{Kofu}
\country{Japan}
}

\author{Jin Cui}
\email{g22dtsa5@yamanashi.ac.jp}
\affiliation{%
  \institution{University of Yamanashi}
  \orcid{0000-0001-9575-3678}
\city{Kofu}
\country{Japan}
}

\author{Yoshimi Suzuki}
\email{ysuzuki@yamanashi.ac.jp}
\affiliation{%
  \institution{University of Yamanashi}
  \orcid{0000-0001-5466-7351}
\city{Kofu}
\country{Japan}
}

\author{Jiyi Li}
\email{jyli@yamanashi.ac.jp}
\affiliation{%
  \institution{University of Yamanashi}
  \orcid{0000-0003-4997-3850}
\city{Kofu}
\country{Japan}
}

\author{Dongjin Yu}
\email{yudj@hdu.edu.cn}
\affiliation{%
  \institution{Hangzhou Dianzi University}
  \orcid{0000-0001-8919-1613}
\city{Hangzhou}
\country{China}
}

\copyrightyear{2024}
\acmYear{2024}
\setcopyright{acmlicensed}\acmConference[SIGIR '24]{Proceedings of the 47th International ACM SIGIR Conference on Research and Development in Information Retrieval}{July 14--18, 2024}{Washington, DC, USA}
\acmBooktitle{Proceedings of the 47th International ACM SIGIR Conference on Research and Development in Information Retrieval (SIGIR '24), July 14--18, 2024, Washington, DC, USA}
\acmDOI{10.1145/3626772.3657799}
\acmISBN{979-8-4007-0431-4/24/07}
\acmPrice{}
\renewcommand{\shortauthors}{Xinfeng Wang et al.}


\begin{abstract}

\input{contents/Abstract.tex}

\end{abstract}

\ccsdesc[500]{Information systems~Recommender systems}

\keywords{Recommendation, Hypergraph Convolution, Over-Smoothing Issue, Individual Bias, Debiasing}

\maketitle

\section{Introduction}
\input{contents/Introduction}

\section{Related Work}

\input{contents/RelatedWork}

\section{Preliminaries}
\input{contents/Preliminaries}

\section{Methodology}
\input{contents/Methodology}

\section{Experiment}
\input{contents/Experiment.tex}

\section{Conclusion}
\input{contents/Conclusion.tex}

\section*{Acknowledgements}
\input{contents/Acknowledgements}


\normalem
\bibliographystyle{ACM-Reference-Format}
\bibliography{sample-base}

\end{document}

%% file: contents/Abstract.tex
Recommender models aimed at mining users' behavioral patterns have raised great attention as one of the essential applications in daily life. Recent work on graph neural networks (GNNs) or debiasing methods has attained remarkable gains. However, they still suffer from (1) over-smoothing node embeddings caused by recursive convolutions with GNNs, and (2) the skewed distribution of interactions due to popularity and user-individual biases. This paper proposes a contextualized and debiased recommender model (CaDRec). To overcome the over-smoothing issue, we explore a novel hypergraph convolution operator that can select effective neighbors during convolution by introducing both structural context and sequential context. To tackle the skewed distribution, we propose two strategies for disentangling interactions: (1) modeling individual biases to learn unbiased item embeddings, and (2) incorporating item popularity with positional encoding. Moreover, we mathematically show that the imbalance of the gradients to update item embeddings exacerbates the popularity bias, thus adopting regularization and weighting schemes as solutions. Extensive experiments on four datasets demonstrate the superiority of the CaDRec against state-of-the-art (SOTA) methods. Our source code and data are released at \href{https://github.com/WangXFng/CaDRec}{https://github.com/WangXFng/CaDRec}.

%% file: contents/Introduction.tex
\begin{figure}
  \includegraphics[width=\linewidth]{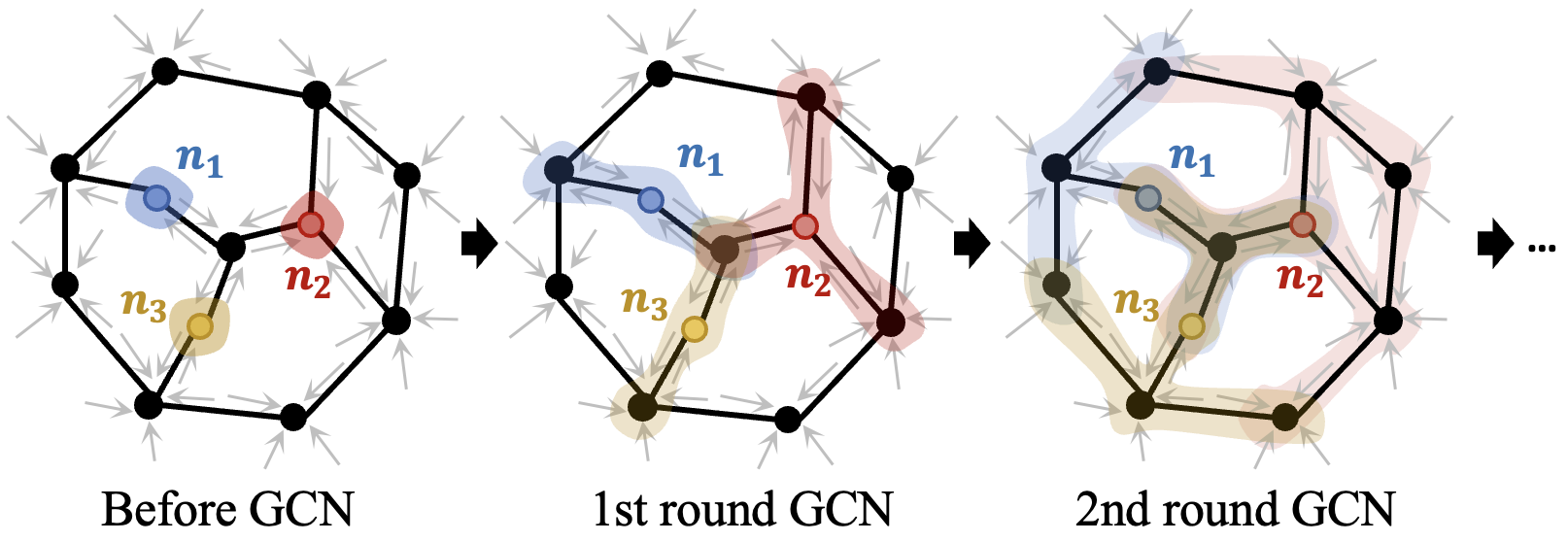}
  \caption{Illustration of the over-smoothing issue. After two rounds of GCNs, the features of $n_1$, $n_2$, and $n_3$ have propagated throughout the graph, making nodes more indistinguishable. }
  \label{fig:motivation}
\end{figure}  

With information overload via Internet services, recommender systems have become inevitable applications in daily life. A reliable recommender system should provide high-quality results by mining behavioral patterns from limited users' interactive histories. One typical approach is to leverage graph convolution networks (GCNs) to propagate node embeddings through user\textendash item interaction edges to capture collaborative signals \cite{he2020lightgcn, hao2021pre, yu2022low}, e.g., self-supervised graph learning ~\cite{yu2021socially}, and hypergraph convolutions ~\cite{xia2022hypergraph, xia2022self, yan2023spatio}. 
However, they still suffer from over-smoothing issues caused by iterative convolutions over a sparse user\textendash item interactive graph based on a static adjacency matrix. As depicted in Fig. \ref{fig:motivation}, given a subgraph with the maximum shortest path of 4, after two rounds of GCN operations, the features of nodes $n_1$, $n_2$, and $n_3$ have propagated across the entire subgraph, resulting in plenty of nodes having similar information and being overly uniform (smooth), as highlighted by blue, red, and yellow edges.
To address the issue, several attempts including users' common interests mining \cite{liu2021interest}, graph structure augmentations \cite{wu2021self}, and node augmentations ~\cite{yu2022graph, yu2023xsimgcl, lin2022improving, cai2022lightgcl}, have been made to increase the dispersion of node embeddings to capture discernible collaborative signals through GCNs. 

However, one major drawback of GCN operators in the approaches mentioned above is that they cannot capture the sequential dependencies of neighbors but only the connectivity between two nodes, even though sequential contexts are an important clue to mitigate over-smoothing issues. For instance, once a user completes the initial volume of a book or movie, it's quite natural that the system would likely suggest its sequel. As shown in Fig. \ref{fig:motivation}, if node $n_3$ refers to a random neighbor and $n_2$ serves as a sequel of $n_1$, a user who has finished experiencing $n_1$ is more likely to prefer $n_2$ than $n_3$. 
This suggests that $n_1$ has a more contextual relationship with $n_2$ in sequential dependencies, compared to $n_3$ in the GCN graph.
The attention mechanism is one solution to capture the sequential dependencies. Several algorithms including graph attention network (GAN)-based techniques \cite{wu2019pd, chen2022multi} have been proposed to select representative items from users' massive sequential interactions that can reflect their preferences \cite{tao2022sminet, wang2023statrl}. More recently, ~\cite{ xia2022hypergraph, wang2023eedn} incorporated transformers with GCNs. All of the approaches gained inspiring results. However, they select ineffective items in the way of modeling user\textendash item interaction as they ignore that both structural and sequential contexts help derive more effective node embeddings during graph convolution.

Another issue is skewed interaction distribution, in which the user\textendash item interaction susceptibly falls into due to diversified biases~\cite{chen2021bias}, e.g., popularity bias, resulting in deviating user representations from their true preferences. 
One attempt is to disentangle the real properties of items from the popularity impacts~\cite{ren2022mitigating, he2022causpref, cao2022disencdr, zhang2023invariant}. It employs two independent neural encoders to tokenize the properties and popularity of items. 
However, they ensure only the separation of popularity and item representations. Such neural encoders might have learned other factors (e.g., gender and age in fairness-aware recommendations \cite{li2021towards, wang2023survey}) instead of item popularity.
Moreover, most of the disentangling approaches do not pay attention to user individual biases. For instance, given the objective scores of (4, 5, 3, 3) which indicate real properties of four items, if an optimistic user gives over-high rating scores, (5, 5, 4, 5), and a serious user leaves critical reviews and rate them with the scores of (3, 5, 1, 2), the individual biases of these two users are (+1, 0, +1, +2) and (-1, 0, -2, -1), respectively, resulting in harmful to understanding the semantic properties of items correctly. Therefore, disentangling the popularity and user individual biases is essential for mining users’ true preferences and item real properties.


In this paper, we propose a contextualized and debiased representation learning network (CaDRec). Instead of simply employing stacked Transformers behind hypergraph convolutions (HGCs) \cite{xia2022self, wang2023eedn}, our approach involves injecting the self-attention (SA) correlation, which serves as a trainable perturbation \cite{goodfellow2014explaining, yu2023xsimgcl} on edges representing sequential relations among nodes, into convolution operators. This is motivated by our decomposition of the operations of HGC and SA, revealing that both operations facilitate informative diffusion relying on two correlation matrices of the same size, although they are distributed in entirely different spaces.
Thus, the HGC operators consider both structural and sequential contexts among neighboring nodes to propagate messages, rather than relying solely on a static adjacency matrix, aiming to break through the over-smoothing bottleneck in HGC.

To make debiased recommendations, we propose two novel strategies to disentangle user\textendash item interactions: (1) in addition to popularity bias~\cite{chen2022co, he2022causpref, zhao2022popularity}, we model user individual bias to promote debiased item representations, and (2) the popularity of an item is encoded through its interaction count with positional encoding~\cite{vaswani2017attention}, which is plug-and-play and interpretable, ensuring that the items of similar popularity are closer in the embedding space. 
The CaDRec fits the popularity and user individual biases during training, while it generates a recommended list via debiased representations of users and items in testing. 
The main contributions of our work can be summarized as follows:

(1) To mitigate over-smooth graph embeddings, we propose a novel HGC operator that effectively propagates information during GCN by considering both structural and sequential contexts. 

(2) To overcome the distribution shift caused by popularity and user-individual biases, we propose an interpretable disentanglement method to derive debiased representations.

(3) We verify that the imbalance of the gradients to update item embeddings exacerbates the popularity bias whereby adopting regularization and weighting schemes as solutions.

(4) Extensive experiments on four real-world public datasets demonstrate that the CaDRec outperforms SOTA methods with the efficiency of time complexity.

%% file: contents/RelatedWork.tex
\noindent
\textbf{Recommendation Approaches for Over-Smoothing Issue.}
The GCN-based methods are prone to the over-smoothing issue, i.e. node embeddings are similar and difficult to discriminate from each other \cite{keriven2022not}.
To mitigate the issue, \citet{liu2021interest} have attempted to utilize the common interests between high-order adjacent users to enhance embedding learning in GCN. Several attempts \cite{yu2021socially, yu2021self, wu2021self, peng2022less, wang2022towards} have explored self-supervised learning techniques to boost the robustness of node embeddings.
 Contrastive augmentations also attracted intensive attention for enhancing the dispersion of node embeddings via contrastive views, such as nodes with their structural neighbors~\cite{lin2022improving}, node embeddings with and without uniform noise~\cite{yu2022graph, yu2023xsimgcl}, and SVD-guided contrastive pairs \cite{cai2022lightgcl}. Likewise, ~\citet{wang2023eedn} mine implicit features to make node embeddings more distinctive.
However, most of the approaches overlook that sequential contexts give a dynamic sequential diversity which boosts graph-based contexts to select effective neighbors during graph convolutions.

\noindent
\textbf{Recommendation Approaches for Debiasing.}
Recommender models are vulnerable to several biases~\cite{chen2021bias}, such as popularity and exposure biases, which result in severe performance drops. To generalize popularity, many works independentize representations of popularity and item property via orthogonal constraints, such as cosine similarity~\cite{ren2022mitigating, zhang2023invariant}, inner product~\cite{chen2022co}, curriculum learning~\cite{zheng2021disentangling}, information theory~\cite{cao2022disencdr}, and anti-preference negative sampling \cite{he2022causpref}. 
They utilize two neural encoders for the separation of popularity and item representations.
However, these encoders might have learned factors (e.g., gender and age) other than popularity.  
%
Besides, several theoretical tools have been explored for debiasing recommendations, such as information bottlenecks \cite{liu2021mitigating, liu2023debiased}, inverse propensity scores~\cite{ding2022addressing, zhang2023recommendation}, upper bound minimization \cite{xiao2022representation}, 
bias-aware contrastive learning~\cite{zhang2022incorporating}, causal analysis \cite{li2023causal}, and quantitative metric \cite{zhou2023adaptive}. \citet{qiu2022contrastive} and \citet{zhang2024mitigating} optimize representation distributions for debaising by counteracting dimensional collapse and embedding degeneration.
Most of the above algorithms overlook that the item representations learned from the data distribution are impeded by user-individual bias. Our decomposition technique is motivated by ~\citet{zheng2021disentangling}. The difference is that we incorporate the individual bias and encode popularity with positional encoding.

%% file: contents/Preliminaries.tex
\subsection{Definitions}

\noindent
\textbf{(User Individual Bias)}.
$\mathbf{e}_u^{indi}$ represents the individual bias of user $u \in \mathcal{U}$ ($|\mathcal{U}|=M$), $\mathbf{E}^{indi} \in
\mathbb{R}^{M \times d_m}$ denotes an embedding matrix, and $d_{m}$ refers to the dimension size.


\noindent
\textbf{(Item Popularity)}.
The popularity embedding $\mathbf{e}_i^{pop}$ of item $i$ is encoded with its global activeness (i.e., its count of interactions, $z_c$) by using the positional encoding, where the element at index $j$ is $\mathrm{cos}(z_c/10000^{(j-1)/d_m})$ for odd $j$ and $\mathrm{sin}(z_c/10000^{j/d_m})$ for even $l$.


\noindent
\textbf{(Item Recommendation)}.
Given a user set $\mathcal{U}$, an item set $\mathcal{I}$, and a user historical interaction set $\mathcal{S}$, where $\mathcal{S}^{(u)} = \{i_{j}\}_{j=1}^{L}$ denotes the item sequence of user $u$, and $L$ refers to the number of items, the goal is to generate a list of the top-$k$ candidate items for user $u$.

\subsection{Perturbation for Modeling Individual Bias}

\begin{figure}
  \includegraphics[width=\linewidth]{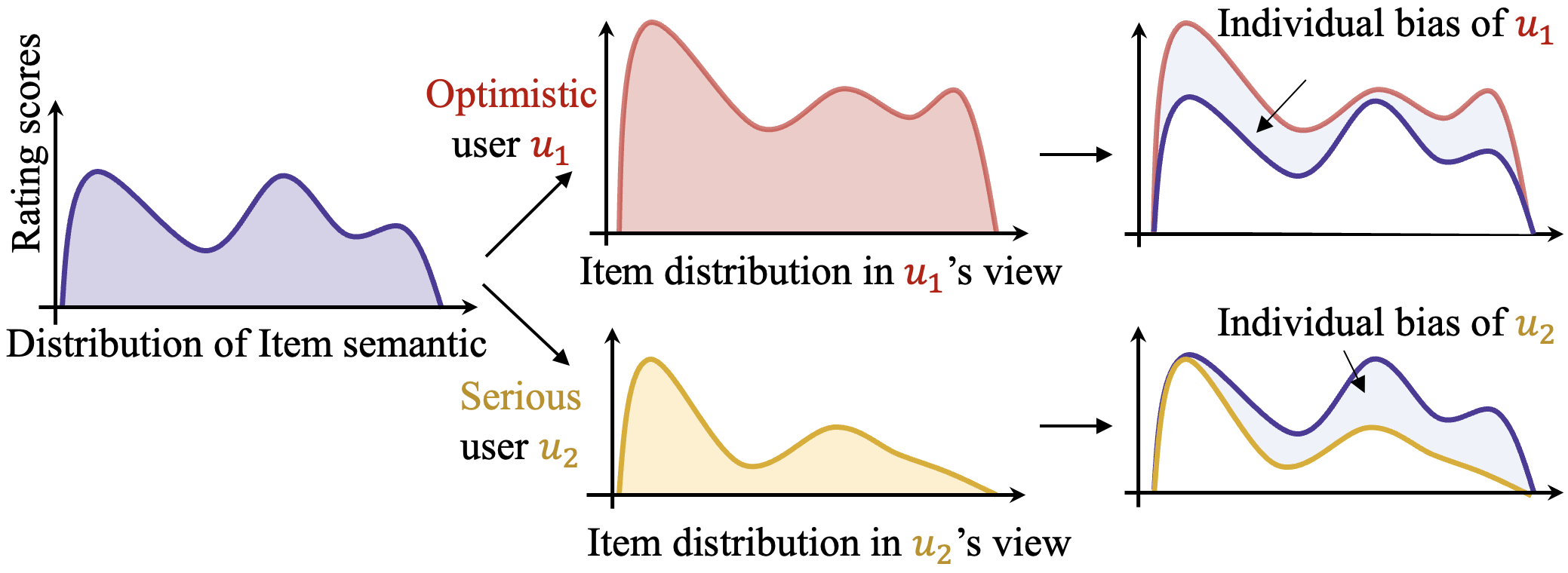}
  \caption{Illustration of Individual Bias.}
  \label{fig:individual_bias}
\end{figure}  

It is often the case that observational interaction distributions are skewed due to diversified latent factors, such as popularity and exposure bias~\cite{chen2021bias}. These latent factors are well-studied in recent years \cite{ren2022mitigating, zhang2023invariant, xiao2022representation, zhang2023recommendation}.
However, they do not mention that optimistic users often give high rating scores while serious people may not, even for the same items, which will prevent capturing semantic properties of preference correctly. 
In Fig.~\ref{fig:individual_bias}, the distribution of item real semantics about its objective rating score is distinct (depicted in purple). However, real-world situations often introduce a user-individual bias (depicted in grey) in this distribution between the item real semantics distribution (in purple color) and the observed item distributions (in red and yellow colors) from users' interactive behaviors. This bias is harmful to understanding items' real semantics for recommender algorithms.



These observations indicate that the observed item distribution in each user's perspective is determined by the item semantic distribution and the user-individual bias. We thus assume that individual bias is a learnable perturbation on item representations when simulating the user-item interactions via graph convolutions. 
Perturbation learning is widely utilized to bolster the robustness and generalization capabilities of models, such as word-level enhancement \cite{huang2022word}, image noise injection \cite{song2023deep}, and node embedding augmentation in recommendation tasks \cite{yu2022graph, yu2023xsimgcl}. Therefore, by urging the learnable perturbation to converge to user-individual bias, we can obtain a robust item semantic distribution that is free from being entangled in individual biases. 

It is noteworthy that user-individual bias as an important clue of user preferences is leveraged to generate user representations in both training and test stages.

%% file: contents/Methodology.tex
\begin{figure*}[ht]
\center{
\begin{minipage}{\linewidth}
\includegraphics[width=\linewidth]{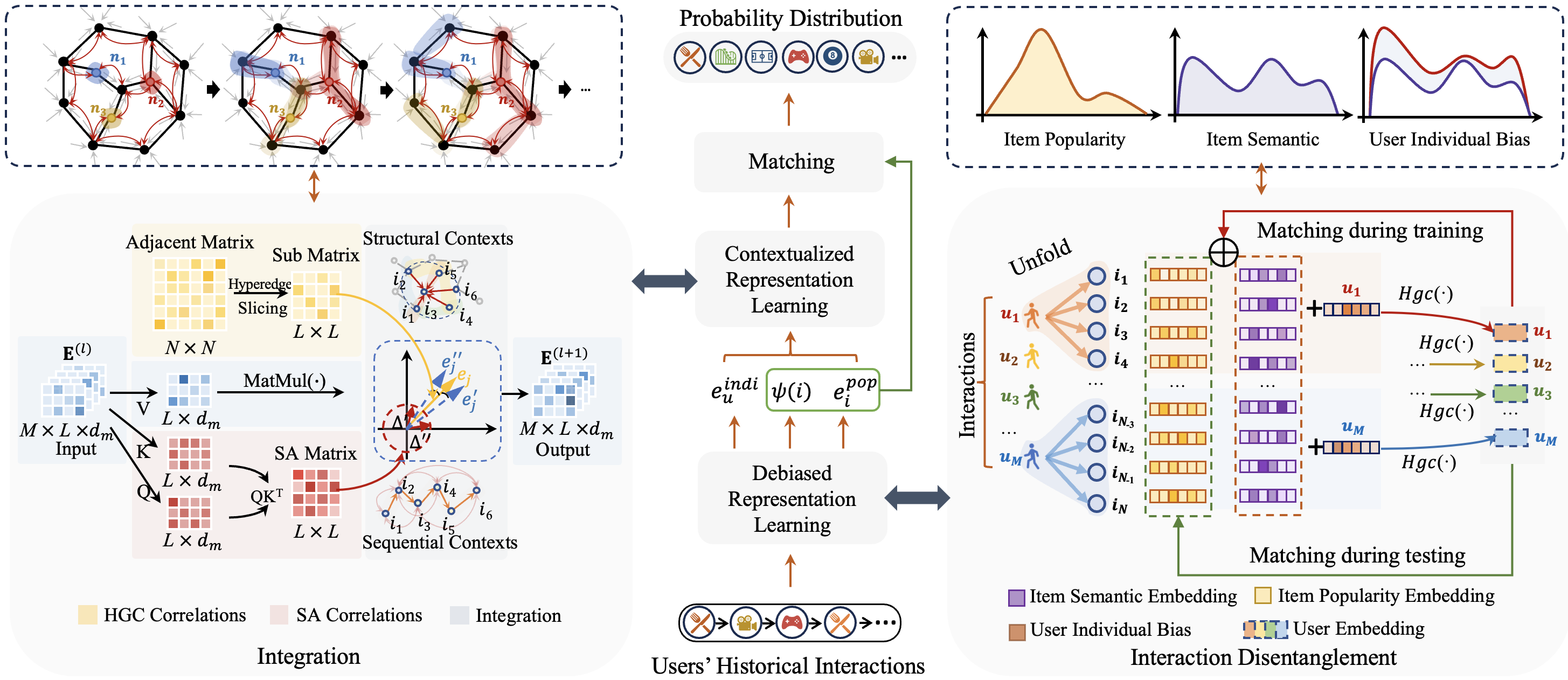}
  \caption{Illustration of the overall framework, consisting of contextualized and debiased representation learning modules. } 
  \label{fig:framework}
\end{minipage}
}
\end{figure*}

 Fig.~\ref{fig:framework} illustrates the framework of CaDRec, consisting of contextualized representation learning and debiased representation learning.

\subsection{Model Basics}
The CaDRec adopts a dot-product model as its backbone, which estimates the preference $\hat{\mathbf{r}}_{u,i} \in \mathbb{R}$
of a user $u \in \mathcal{U}$ for an item $i \in \mathcal{I}$ by an inner-product between
the embeddings of user $u$ and item $i$:
\begin{equation}
\hat{\mathbf{r}}_{u,i} = <\phi(u), \psi(i)>,\\
\end{equation}

\noindent
where $\phi(u) \in \mathbb{R}$ and $\psi(i) \in \mathbb{R}$ are the embeddings of $u$ and $i$, respectively. $\phi:\mathcal{U} \xrightarrow{} \mathbb{R}^{d_m}$ and $\psi:\mathcal{I} \xrightarrow{} \mathbb{R}^{d_m}$ denote the feature mappings of the model. 
We initiate item embeddings by the uniform distribution and exert users' historical interactions $\mathcal{S}^{(u)}$ to generate user embeddings via $f(\cdot)$, which are given by:
\begin{equation}
\psi(i) \sim Uniform(-1, 1), \phi(u) \sim f(\{\psi(i_k) : i_k \in \mathcal{S}^{(u)}\}). \\
\end{equation}

\subsection{Contextualized Representation Learning}
We first introduce the hypergraph convolution (HGC), the relation between HGC and Transformer, and their incorporation. We then show the process of learning users' contextualized representations. 

\noindent
\textbf{Graph Convolution (GCN).} Following the works by~\cite{he2020lightgcn, wang2023eedn}, we removed the global weight matrix and added a linear feature transformation ($\mathbf{W}_1 \in \mathbb{R}^{d_m \times d_m}$). The matrix equivalence form of informative diffusion in the GCN is as follows:
\begin{equation}
\mathbf{E}^{(l+1)}= \hat{\mathbf{A}} \alpha_1(\mathbf{E}^{(l)} \mathbf{W}_1), \\
\end{equation}

\noindent
where $\mathbf{E}^{(l)} \in \mathbb{R}^{N \times d_m}$ indicates the output of the $l$-th GCN layer, $\alpha_1$ is the ELU activation function, and $\hat{\mathbf{A}}$ denotes the symmetrically normalized adjacency matrix:
\begin{equation}
\hat{\mathbf{A}} = \mathbf{D}^{-1/2} \mathbf{A} \mathbf{D}^{-1/2},  \\
\end{equation}

\noindent
where $\mathbf{A} \in \mathbb{R}^{N \times N}$ denotes the adjacency matrix, in which $a_{jk} = 1$, if the same user has interacted with item $i_j$ and item $i_k$, otherwise 0. $\mathbf{D} \in \mathbb{R}^{N \times N}$ refers to the diagonal matrix in which each entry $d_{jk}$ indicates the number of nonzero entries in the $j$-th row of $\mathbf{A}$.

\noindent
\textbf{Hypergraph Convolution (HGC).} 
Inspired by the EEDN \cite{wang2023eedn}, we utilized a user hypergraph to enhance signal propagation during GCN via learning local contexts from user hyperedges. 
Note that the multiplication of $\mathbf{E}^{(l)}$ and a hyperedge $\mathcal{H}^{(u)}$ can be implemented by the slicing operation that selects corresponding rows in $\mathbf{E}^{(l)}$. Thus, $\hat{\mathbf{A}}$ and $\mathbf{E}^{(l)}$ are substituted by $\hat{\mathbf{A}}[\mathcal{S}^{(u)}] \in \mathbb{R}^{L \times L}$, and $\mathbf{E}^{(l)}[\mathcal{S}^{(u)}] \in \mathbb{R}^{L \times d_m}$, respectively:
\begin{equation}
\begin{aligned}
\hat{\mathbf{A}} \ \ &\gets \ \hat{\mathbf{A}}\ [\mathcal{S}^{(u)}] \ =\mathcal{H}^{(u)} \hat{\mathbf{A}} {\mathcal{H}^{(u)}}^\top, \\
\mathbf{E}^{(l)} &\gets \mathbf{E}^{(l)}[\mathcal{S}^{(u)}] = \mathbf{E}^{(l)} {\mathcal{H}^{(u)}}^\top,\\
\end{aligned}
\end{equation}
\noindent
where $[\cdot]$ refers to the slicing operation and $\mathcal{H}^{(u)}$ indicates the hyperedge of the user $u$. As the slicing operation takes $O(1)$, the computational complexity of HGC reduces from $O(N^2)$ to $O(L^2)$, where $N \gg L$. Nevertheless, the HGC cannot handle the over-smoothing issue caused by iterative message-passing over the static structural graph. 

\noindent
\textbf{Integration of HGC and SA.} 
Indeed, 
there exists a strong underlying relation between HGC and Transformer, in the sense that these two paradigms aggregated the input signals with weight by two correlation matrices of the same size. Specifically, the sub adjacency matrix $\hat{\mathbf{A}}$ $\in \mathbb{R}^{L \times L}$ refers to the structural relationships between items within the user's hyperedge, while the SA matrix $\mathbf{Q}\mathbf{K}^\top \in \mathbb{R}^{L \times L}$ learns the sequential dependencies between the user's historical interactions. However, these relations are distributed from completely different spaces. 
Inspired by the random perturbation \cite{goodfellow2014explaining, yu2023xsimgcl}, we exert the SA relation as a perturbation $\Delta'$ to aid the GCN operator in selecting effective neighbors, as the integration module depicted in Fig.~\ref{fig:framework}.
Thus, informative diffusion is given by:
\begin{equation}
\label{eq:hgc}
\begin{aligned}
&\hat{\mathbf{E}}^{(l)}= (\hat{\mathbf{A}} + \Delta')\alpha_1(\mathbf{V}) , \quad  \Delta' =  \delta \cdot \mathrm{Norm}(\frac{\mathbf{Q}\mathbf{K}^\top}{\sqrt{d_m}}), \\
&\mathbf{Q}                                                                                                             = \mathbf{E}^{(l)}\mathbf{W}^Q, \quad \mathbf{K}  =  \mathbf{E}^{(l)}\mathbf{W}^K, \quad \mathbf{V} = \mathbf{E}^{(l)}\mathbf{W}_1,
\end{aligned}
\end{equation}

\noindent
where ${\mathbf W}^Q  \in \mathbb{R}^{d_m \times d_m}$ and ${\mathbf W}^K \in \mathbb{R}^{d_m \times d_m}$ are weight matrices, $\mathrm{Norm}(\cdot)$ indicates L$_2$ normalization, and $\delta$ is a hyperparameter. As such, HGC can promote diverse representations by considering both structural and sequential contexts, with a time complexity reduced from $O(2L^2)$ to $O(L^2$+$L)$. The multi-head (MH) block is used to learn features in multiple subspaces. 
The $\mathbf{E}^{(l+1)}$ is given by the following equation:
\begin{equation}
\mathbf{E}^{(l+1)} = \mbox{MultiHead}(\hat{\mathbf{E}}^{(l)}_1,\hat{\mathbf{E}}^{(l)}_2,...,\hat{\mathbf{E}}^{(l)}_{z_h}),  \\
\end{equation}
where $z_h$ denotes the number of heads. A simple summation as the aggregation function works better than a feed-forward network. 

Consider the item embeddings as the input $\mathbf{E}^{(0)}$ of the first HGC layer, 
and we obtain user $u$'s embedding by feeding the output $\mathbf{h}$ of the last HGC layer into an average-pooling layer and L$_2$ normalization layer as follows:
\begin{equation}
\begin{aligned}
\phi(u) &= \mathrm{Norm}(\mathrm{AvgPooling}(\mathbf{h})), \\
\end{aligned}
\end{equation}

To exploit sequential contexts, we adopt a multi-label cross-entropy loss instead of the widely used Bayesian personalized ranking (BPR) loss~\cite{rendle2012bpr} as the objective function:
\begin{equation}
\label{eq:cross_entropy}
  \mathcal{L}_{rating} = - \sum_{(u \in \mathcal{U})} \gamma_{1} \log(\alpha_2(\hat{\mathbf{r}}_{u,1})) + ... + \gamma_{N} \log(\alpha_2(\hat{\mathbf{r}}_{u,N})),
\end{equation}
\noindent
where $\mathbf{\gamma} \in \mathbb{R}^{N}$ refers to a true label vector, and each dimension of the vector equals 1 when the user has ever visited the item; otherwise, 0. $\alpha_2$ is the $\mathrm{sigmoid}$ activation function.

\subsection{Debiased Representation Learning}

We introduce two debiasing methods, i.e., interaction disentanglement and embedding debiasing to enhance recommendations.


\noindent
\textbf{Interaction Disentanglement.}
As shown in Fig. \ref{fig:framework}, we built three tokens to disentangle interactions: user individual bias, item popularity, and item real semantic. 
%
To disentangle item representations from user individual bias, similar to Eq. (\ref{eq:hgc}), we regard $\mathbf{e}^{indi}_u$ as a learnable perturbation $\Delta''$ to the items' real semantics to fitting biased interactions. The feature vector $\phi(u)$ of user $u$ is given by:
\begin{equation}
\label{eq:individual_bias_learning}
\begin{aligned}
\phi(u) &= Hgc(\{\psi(i_k) + \Delta'':{i_k}\in \mathcal{H}^{(u)} \}), \\
                    \Delta'' &= \mathrm{sign}(\psi(i_k))\odot \mathrm{Norm}(\mathbf{e}^{indi}_u), 
\end{aligned}
\end{equation}

\noindent
where $Hgc(\cdot)$ denotes the representation of the HGC with injecting SA. $\mathbf{e}^{indi}_u \in \mathbb{R}^{1 \times d_m}$ learns the individual bias of user $u$ such that $\psi(i_k)$ has less individual bias, and $\odot$ is a broadcast multiplication. 


 
We assume that the user embedding $\phi(u)$ has little popularity. Therefore, we do not take into account $\bar{\textbf{e}}^{pop}_u = \mathrm{AvgPooling}(\{\mathbf{e}^{pop}_{i_k}, i_k \in \mathcal{S}^{(u)}\})$ and $\textbf{e}^{pop}_{i}$ for testing. We refined the rating score as follows:
\begin{equation}
\hat{\textbf{r}}_{u,i} =
    \begin{cases}
    <[\phi(u), \beta_1\bar{\textbf{e}}^{pop}_u], [\psi(i), \beta_1\textbf{e}^{pop}_{i}]>, & \mathrm{during\ training} \\
    <\phi(u), \psi(i)>, & \mathrm{otherwise} \\
    \end{cases}
\end{equation}

\noindent
where $[\cdot, \cdot]$ denotes a concatenation and $\beta_1$ is a hyperparameter that varies the influence level of popularity.
Here, $\phi(u)$ denotes the pure preference of user $u$, and the real semantic $\psi(i)$ of item $i$ is free from being entangled in popularity and user-individual biases. 

\noindent
\textbf{Debiasing on Embeddings.}
We revisit the gradient of the rating loss for the embeddings $\Theta$ (i.e., user or item embeddings) as follows:
\begin{equation}
\label{eq:gradient}
\begin{aligned}
\frac{\partial\ \mathcal{L}_{rating}}{\partial\ \Theta} &=\sum_{(u \in \mathcal{U})} \frac{\partial\ -\mathbf{\gamma}^{\top}\mathrm{log} \alpha_2(\mathcal{R})}{\partial\ \Theta}, \\
&=\sum_{(u \in \mathcal{U})} - \mathbf{\gamma}^{\top} (1+e^{-\mathcal{R}}) \frac{-e^{-\mathcal{R}} \cdot (-1)} {(1+e^{-\mathcal{R}})^2}\cdot \frac{\partial\ \mathcal{R}}{\partial\ \Theta}, \\
&\varpropto \frac{-\mathbf{\gamma}^{\top} e^{-\mathcal{R}}} {1+e^{-\mathcal{R}}}\cdot \frac{\partial\ \mathcal{R}}{\partial\ \Theta}.
\end{aligned}
\end{equation}

 We further define the user embeddings as $\mathcal{P}$, and the embeddings of items that users have \underline{\textbf{I}}nter\underline{\textbf{A}}cted (IA) and \underline{\textbf{F}}uture-\underline{\textbf{I}}nter\underline{\textbf{A}}cted (FIA) as $\mathcal{Q}^{IA}$ and $\mathcal{Q}^{FIA}$, respectively. Note that (1) a user's embedding is learned from the user's IA items through $Hgc(\cdot)$, (2) negative samples have been filtered out by $\mathbb{\gamma}^\top$, and (3) $\frac{-\mathbf{\gamma}^\top e^{-\mathcal{R}}} {1+e^{-\mathcal{R}}}$ is a constant matrix for $\frac{\partial\ \mathcal{R}}{\partial\ \Theta}$. We thus derive $\mathcal{R} = \mathcal{P} \cdot \mathcal{Q}^\top$, $\mathcal{P} = Hgc(\mathcal{Q}^{IA})$, and $\mathcal{Q} = [\mathcal{Q}^{IA},\mathcal{Q}^{FIA}]$. To sum up, the gradients for updating embeddings are as follows:
\begin{equation}
\begin{aligned}
\frac{\partial\ \mathcal{R}}{\partial\ \Theta}\! &= \!\frac{\partial\ \mathcal{P} \cdot \mathcal{Q}^\top}{\partial\ \Theta} 
= \!\frac{\partial\ Hgc(\mathcal{Q}^{IA}) \cdot [\mathcal{Q}^{IA},\mathcal{Q}^{FIA}]^\top}{\partial\ \Theta},\\
    &=\begin{cases}
    Hgc(\mathcal{Q}^{IA})\!+\!Hgc'(\mathcal{Q}^{IA}) [\mathcal{Q}^{IA},\mathcal{Q}^{FIA}]^\top,&\Theta\!=\!\mathcal{Q}^{IA},  \\
    Hgc(\mathcal{Q}^{IA}), &\Theta\!=\!\mathcal{Q}^{FIA},  \\
    0, & \mathrm{otherwise}. \\
    \end{cases}
\end{aligned}
\end{equation}

\begin{figure}
  \includegraphics[width=\linewidth]{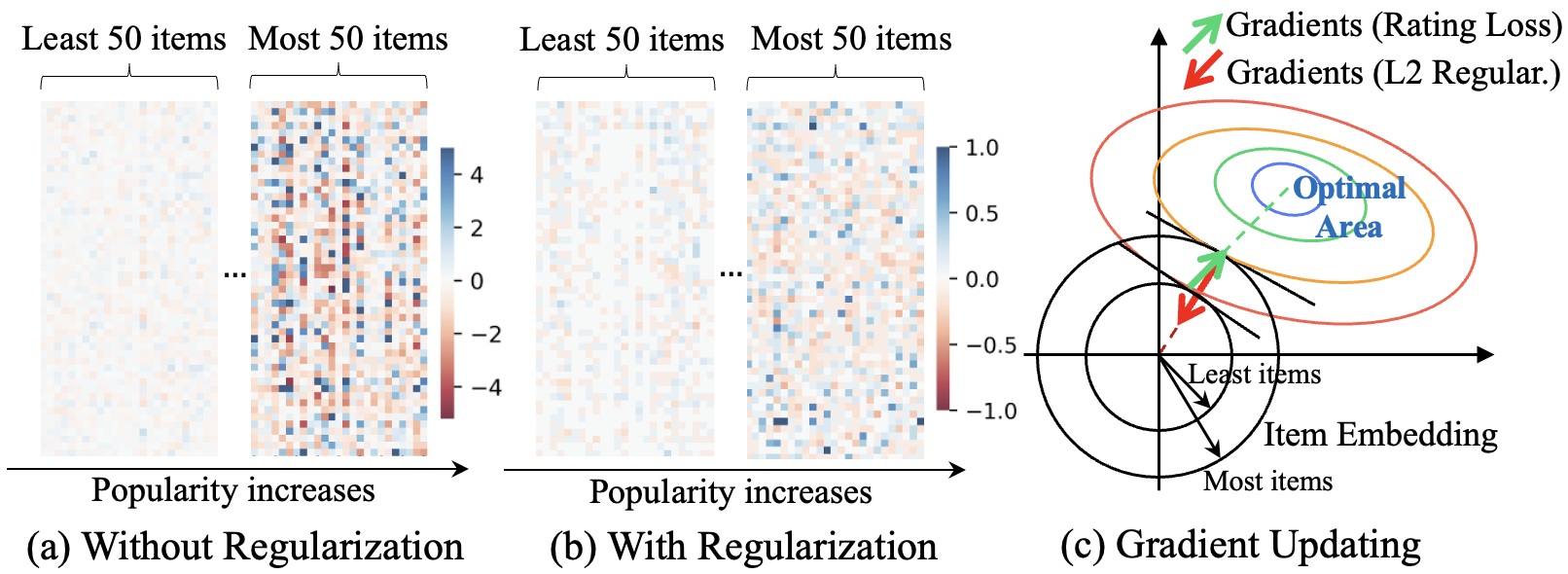}
  \caption{Illustration of item embedding regularization. (a) and (b) denote the well-trained item embeddings without and with regularization in the Yelp2018 dataset, respectively. }
  \label{fig:normalization}
\end{figure}

Through the derivation, we observed two issues in the vanilla multi-label cross-entropy function. The first issue is that there are no penalties for negative samples, resulting from a $\gamma_i$ of 0 for a negative sample $i$, while a positive sample could be a negative sample for another user. This makes the popular items overly encouraged. In Fig.~\ref{fig:normalization} (a), we can see that the variance of embedding vectors obtained by the 50 most popular items was significantly larger than that of the 50 least popular items. This indicates that the value of the inner products of the feature vectors of any user and popular items is likely to be a larger value than those of unpopular ones, resulting in the model recommending the most popular items over users' pure preferences. More specifically, the blue circle in Fig.~\ref{fig:normalization} (c) indicates the optimal area for item embeddings on the training set, while they are overfitting caused by the popularity bias. To address this issue, we employed L$_2$ regularization to the positive samples (i.e., the IA items) during the training stage. 
Formally, given the indicator function $I_{(\cdot)}$, the learning rate $\eta$, the hyperparameter of $\beta_2$ to control the penalization level, and the user set $\mathcal{U}^B$ in the training batch, the gradients to update $\psi(i)$ are formulated as follows:
%
%
\begin{small}
\begin{equation}
\label{eq:nomalization}
\begin{aligned}
\psi(i) \xleftarrow{} \underbrace{(1-\sum_{u \in \mathcal{U}^B} I_{(i \in \mathcal{S}^{(u)})} \cdot \eta\beta_2) \cdot \psi(i)}_{\mathrm{Regularization\ term}} - \underbrace{\sum_{u \in \mathcal{U}^B}I_{(i \in \mathcal{S}^{(u)})} \cdot \eta \frac{\partial\ \mathcal{L}_{rating}}{\partial\ \psi(i)}}_{\mathrm{Rating\ loss\ term}}.
\end{aligned}
\end{equation}
\end{small}

As such, the rating loss term enables item embeddings to learn real semantic, i.e., embeddings are updated in the correct direction, as illustrated in the green arrow of Fig.~\ref{fig:normalization} (c). Simultaneously, the regularization term prevents popular items from being overweighted by reducing the embedding complexity, which is shown in the red arrow. As illustrated in (a) and (b), we empirically showed significant mitigation of gradient imbalance between the most and least popular item embeddings after regularization.

The second issue is that the gradients for updating IA and FIA item embeddings are often imbalanced as there is always one more term, i.e., $Hgc'(\mathcal{Q}^{IA}) [\mathcal{Q}^{IA},\mathcal{Q}^{FIA}]^\top$, for the IA items.
Thus, we utilize a weighting scheme \cite{ma2018point, wang2023eedn} to balance the gradients. The final loss function is given by:
\begin{equation}
\begin{scriptsize}
  \mathcal{L}_{total} = 
  \sum_{(u \in \mathcal{U})}(-\mathbf{C}_{u,*}\odot\mathbb{\gamma}^{\top} \log(\mathrm{\alpha_2(\mathbf{r}_{u,*})}) + \beta_2\sum_{(i \in \mathcal{S}^{(u)})} ||\psi(i)||_2 ),
\end{scriptsize}
\end{equation}

\noindent
where $\odot$ denotes the element-wise multiplication and $||\cdot||_2$ refers to L$_2$ regularization. $\mathbf{c}_{u,i}$ is given by:
\begin{equation}
\begin{aligned}
\begin{small}
\mathbf{c}_{u,i} = 
    \begin{cases}
    \lambda_1 & \mathrm{user}\  u\  \mathrm{interacted\ with\ item} \ i, \\
    \lambda_2 & \mathrm{user}\  u \ \mathrm{will\  interact\  with\ item\ } i, \\
    0 & \mathrm{otherwise},  \\
    \end{cases}
\end{small}
\end{aligned}
\end{equation}
\noindent
where $\lambda_1$ and $\lambda_2$ are two hyperparameters used to balance the weights between the IA and FIA items for users.

%% file: contents/Experiment.tex

\begin{table}
\center
\small
  \caption{\small{Statistics of the datasets. ``\#Avg." denotes the average count of users' interactions.}}
  \label{tab:dataset}
\resizebox{.94\columnwidth}{!}{ 
  \begin{tabular}{c|r|r|r|c|c}  
\toprule
    {Dataset} &{\#User} &{\#Item} &{\#Interaction}& {\#Avg.} & Density\\
\midrule
    {Yelp2018} &31,668& 38,048& 1,561,406& 49.3 & 0.13\%  \\
    {Foursquare} &7,642& 28,483& 512,532& 68.2 & 0.24\%   \\
    {Douban-book} &12,859& 22,294& 598,420& 46.5 & 0.21\%  \\
    {ML-1M} &6,038& 3,533& 575,281& 95.3 & 2.7\%  \\
\bottomrule
\end{tabular}
}
\end{table}

\begin{table*}
  \caption{Performance comparison of different recommendation models. \textbf{Bold}: best, \underline{underlined}: second best.}
  \label{tab:comparison}
\resizebox{1\linewidth}{!}{ 
  \begin{tabular}{c|cc|cc|c|ccc|ccccccc|c}
      \toprule
     \multirow {2}{*}{Metric } &\multicolumn{2}{c|}{GCN-based} 
    &\multicolumn{2}{c|}{SA-based} & \small{Diff-based} & \multicolumn {3}{c|}{Debias-\small{based} } & \multicolumn {7}{c|}{Over-smoothing alleviation (OSA)-based } & \textbf{CaDRec}
    \\ 
& \footnotesize{LightGCN} & {LCFN} & \footnotesize{AutoCF} & \footnotesize{STaTRL} & \small{DiffRec} & {DICE} & \small{InvCF} & \footnotesize{LightGCL} & {SGL}  & \small{IMP}\footnotesize{GCN} & GDE & {HCCF} & {NCL}  & \footnotesize{XSimGCL} & {EEDN}& \textbf(ours) \\
\midrule
    \multicolumn{17}{c}{Yelp2018} \\
\midrule
R@5 & 0.0191 & 0.0209 & 0.0221& 0.0215& 0.0236& 0.0204 & 0.0213 & 0.0195 & 0.0220 & 0.0188& 0.0175 & 0.0221& 0.0223& 0.0251& \underline{0.0267} & \textbf{0.0275} \\ 
N@5 & 0.0371 & 0.0385 & 0.0420& 0.0415& 0.0402& 0.0356 & 0.0399 & 0.0393 & 0.0421 & 0.0354& 0.0329 & 0.0425& 0.0428& 0.0478& \underline{0.0508} & \textbf{0.0533} \\
R@10& 0.0347 & 0.0357 & 0.0391& 0.0372& 0.0414& 0.0320 & 0.0377 & 0.0372 & 0.0384 & 0.0293& 0.0280 & 0.0394& 0.0390&  0.0445& \underline{0.0457} & \textbf{0.0474} \\
N@10& 0.0397 & 0.0378 & 0.0447& 0.0420& 0.0471& 0.0401 & 0.0433 & 0.0414 & 0.0447 & 0.0343& 0.0339 & 0.0454& 0.0456&  0.0502& \underline{0.0531} & \textbf{0.0558} \\
R@20& 0.0582 & 0.0564 & 0.0662& 0.0662& 0.0693& 0.0570 & 0.0657 & 0.0674 & 0.0659 & 0.0504& 0.0490 & 0.0688& 0.0665& 0.0745& \underline{0.0760} & \textbf{0.0792} \\
N@20& 0.0484 & 0.0476 & 0.0546& 0.0501& 0.0586& 0.0410 & 0.0543 & 0.0587 & 0.0549 & 0.0400& 0.0375 & 0.0565& 0.0558& 0.0635& \underline{0.0639} & \textbf{0.0667} \\
\midrule
 \multicolumn{17}{c}{Foursquare} \\
\midrule
R@5 & 0.0480 & 0.0506 & 0.0473& 0.0472& 0.0492& 0.0498 & 0.0419 & 0.0487& 0.0471 & 0.0371& 0.0351 & 0.0489 & 0.0511& 0.0459& \underline{0.0554} & \textbf{0.0561} \\
N@5 & 0.0784 & 0.0786 & 0.0743& 0.0763& 0.0773& 0.0702 & 0.0625 & 0.0765& 0.0753 & 0.0607& 0.0571 & 0.0785 & 0.0834& 0.0721& \underline{0.0867} & \textbf{0.0879} \\
R@10& 0.0725 & 0.0731 & 0.0741& 0.0735& 0.0744& 0.0750 & 0.0677 & 0.0719& 0.0728 & 0.0660& 0.0626 & 0.0777 & 0.0788& 0.0727& \underline{0.0852} & \textbf{0.0867} \\
N@10& 0.0795 & 0.0769 & 0.0771& 0.0778& 0.0790& 0.0732 & 0.0681 & 0.0777& 0.0772 & 0.0743& 0.0710 & 0.0759 & 0.0854& 0.0755& \underline{0.0886} & \textbf{0.0898} \\
R@20& 0.1094 & 0.1165 & 0.1121& 0.1106& 0.1102& 0.1010 & 0.1069 & 0.1067& 0.1102 & 0.0968& 0.0943 & 0.1121 & 0.1206& 0.1138& \underline{0.1270} & \textbf{0.1314} \\
N@20& 0.0934 & 0.0924 & 0.0904& 0.0916& 0.0921& 0.0827 & 0.0828 & 0.0920& 0.0914 & 0.0868& 0.0796 & 0.0964 & 0.1012& 0.0905& \underline{0.1047} & \textbf{0.1068} \\
\midrule
    \multicolumn{17}{c}{Douban-book} \\
\midrule
R@5 & 0.0640 & 0.0623 & 0.0615& 0.0693& 0.0650& 0.0667 & 0.0654 & 0.0666& 0.0790 & 0.0551 & 0.0455 & 0.0756 & 0.0930& \underline{0.1019} & 0.1015& \textbf{0.1055} \\ 
N@5 & 0.1170 & 0.1150 & 0.1215& 0.1370& 0.1225& 0.1288 & 0.1192 & 0.1290& 0.1426 & 0.1038 & 0.0996 & 0.1406 & 0.1661& 0.1901& \underline{0.1912} & \textbf{0.2018} \\
R@10& 0.0972 & 0.0992 & 0.0927& 0.0994& 0.0956& 0.1036 & 0.1089 & 0.1145& 0.1189 & 0.0884 & 0.0740 & 0.1168 & 0.1340& 0.1448& \underline{0.1450} & \textbf{0.1504} \\
N@10& 0.1165 & 0.1123 & 0.1175& 0.1286& 0.1180& 0.1221 & 0.1199 & 0.1452& 0.1418 & 0.1056 & 0.1005 & 0.1416 & 0.1625& 0.1817& \underline{0.1822} & \textbf{0.1918} \\
R@20& 0.1455 & 0.1457 & 0.1344& 0.1401& 0.1369& 0.1520 & 0.1410 & 0.1624& 0.1713 & 0.1293 & 0.1146 & 0.1747 & 0.1898& \underline{0.1989} & 0.1954& \textbf{0.2085} \\
N@20& 0.1253 & 0.1368 & 0.1240& 0.1313& 0.1228& 0.1409 & 0.1356 & 0.1501& 0.1514 & 0.1105 & 0.1497 & 0.1585 & 0.1713& \underline{0.1867} & 0.1852& \textbf{0.1960} \\
\midrule
    \multicolumn{17}{c}{ML-1M} \\
\midrule
R@5 & 0.1067 & 0.0879 & 0.0927& 0.0913& 0.0984& 0.0998 & 0.0935 & 0.1025& 0.1111 & 0.1007& 0.0981& 0.0959 & 0.1128& \underline{0.1161} & 0.1075& \textbf{0.1220} \\ 
N@5 & 0.3237 & 0.2653 & 0.2728& 0.2614& 0.2893& 0.2876 & 0.2827 & 0.3210& 0.3304 & 0.2488& 0.2891& 0.2819 & \underline{0.3419} & 0.3379& 0.3052& \textbf{0.3506} \\
R@10& 0.1708 & 0.1450 & 0.1599& 0.1535& 0.1530& 0.1677 & 0.1617 & 0.1687& 0.1751 & 0.1493& 0.1608& 0.1607 & 0.1800& \underline{0.1834} & 0.1675& \textbf{0.1933} \\
N@10& 0.3057 & 0.2450 & 0.2621& 0.2518& 0.2673& 0.2690 & 0.2663 & 0.2997& 0.3110 & 0.2478& 0.2977& 0.2585 & \underline{0.3224} & 0.3210& 0.2723& \textbf{0.3281} \\
R@20& 0.2622 & 0.2358 & 0.2287& 0.2423& 0.2276& 0.2514 & 0.2361 & 0.2517& 0.2622 & 0.2317& 0.2509& 0.2321 & 0.2727& \underline{0.2747} & 0.2665& \textbf{0.2907} \\
N@20& 0.3076 & 0.2556 & 0.2686& 0.2603& 0.2622& 0.2863 & 0.2703 & 0.3002& 0.3120 & 0.2628& 0.2799& 0.2706 & \underline{0.3227}& 0.3203& 0.2927& \textbf{0.3261} \\
    \bottomrule
  \end{tabular}
}
\end{table*}

\subsection{Experimental Setup}
\noindent
\textbf{Datasets.} 
We performed the experiments on four benchmark datasets: Yelp2018 ~\cite{he2020lightgcn, yu2023xsimgcl, wang2023eedn}, Foursquare~\cite{wang2023eedn}, Douban-book~\cite{yu2021socially, yu2023xsimgcl}, and ML-1M~\cite{lin2022improving}. The data statistics are shown in Table \ref{tab:dataset}. Following related work \cite{wang2023eedn, wang2023statrl,yu2023xsimgcl}, we used the earliest 70\% check-ins as training data, the most recent 20\% check-ins as test data, and the remaining 10\% as validation data for each user. For baselines that obtain the best performance in the training and test sets with a ratio of 8:2, we merged our training and validation set as a training set. Their test set is the same as our model. We used two metrics: Recall@$K$ (R@$K$) and NDCG@$K$ (N@$K$) with $K$ $\in$ \{5,10,20\}. 


\noindent
\textbf{Baselines.} We compared our CaDRec with the following fifteen related works as baselines which are divided into five groups: 

\begin{itemize}[nosep,labelindent=0em,leftmargin=1em,font=\normalfont]
\item \textit{Graph Convolution \textbf{(GCN)-based} method}: (1) \textbf{LightGCN}~\cite{he2020lightgcn} simplifies GCN to make it more suitable for recommendation purposes, and (2) \textbf{LCFN}~\cite{yu2022low}  adopts a low-pass collaborative filter paradigm with trainable kernels to make recommendations.

\item \textit{Self-Attention \textbf{(SA)-based} method}: (1) \textbf{AutoCF}~\cite{autocf2023} augments data via self-supervised learning, and (2) \textbf{STaTRL}~\cite{wang2023statrl} explores the spatiotemporal and text information for recommendations.

\item \textit{Diffusion \textbf{(Diff)-based}  methods}: \textbf{DiffRec} \cite{wang2023diffusion} learns the generative process in a denoising manner for recommendations.

\item \textit{Debiasing \textbf{(Debias)-based} methods}: (1) \textbf{DICE}~\cite{zheng2021disentangling} structurally disentangles interest and conformity to recommend items, (2) \textbf{InvCF}~\cite{zhang2023invariant} disentangles representations of preference and popularity semantics, and (3) \textbf{LightGCL}~\cite{cai2022lightgcl} impairs the model's generality and robustness via a light contrastive paradigm; 

\item \textit{Over-smoothing alleviation \textbf{(OSA)-based} methods}: (1) \textbf{SGL}~\cite{wu2021self} reinforces node representations via self-discrimination, (2) \textbf{IMP-GCN}~\cite{liu2021interest} performs GCNs inside subgraphs with users' similar interests, (3) \textbf{GDE}~\cite{peng2022less} replaces neighborhood aggregation with a denoising filter, (4) {\textbf{HCCF}}~\cite{xia2022hypergraph} exploits cross-view contrastive hypergraphs to learn collaborative relations, 
(5) \textbf{NCL}~\cite{lin2022improving} regards users (or items) and their structural neighbors as contrastive pairs, (6) \textbf{XSimGCL}~\cite{yu2023xsimgcl} creates contrastive views to add uniform noise into embedding spaces, and (7) {\textbf{EEDN}}~\cite{wang2023eedn} mines implicit features to learn rich interactive features between users and items. 

\end{itemize}

\noindent
\textbf{Implementation.}
The best hyperparameters of the CaDRec were sampled as follows: $\delta$ is 0.1 for Foursquare and Douban-book, 0.5 for Yelp2018, and 0 for ML-1M. $\beta_1$,
$\beta_2$, $\lambda_1$, and $\lambda_2$ were set to 0.25, 0.6, 2.3, and 7.0 for Yelp2018, 0.7, 0.05, 0.33, and 0.44 for Foursquare, 0.4, 0.06, 0.46 and 0.41 for ML-1M, and 0.07, 0.65, 0.98, and 0.75 for Douban-book, respectively.  $d_m$ was 768 for Foursquare and 512 for the others. The learning rate $\eta$ was 1e-2. The number of HGC layers $z_l$ was 1, and the number of heads $z_h$ was set to 3 and 2 for Foursquare and ML-1M, and 1 for others. These hyperparameters were tuned using Optuna\footnote{https://github.com/pfnet/optuna}. 
The parameters for the baselines were tuned to attain the best performance or set as reported in the original papers. For a fair evaluation, we conducted each experiment five times and obtained the average result. We implemented the CaDRec with Pytorch and experimented on the Nvidia GeForce RTX 3090 (24GB memory).

\subsection{Experimental Results}

We compared our method with SOTA recommender approaches. The results are displayed in Table \ref{tab:comparison}. We can see that CaDRec is consistently superior to all baselines, specifically, the improvements compared with the runner-ups EEDN, NCL, and XSimGCL were 3.7\% $\sim$ 5.3\% on Yelp2018, 1.3\% $\sim$ 3.5\% on Foursquare, 3.6\% $\sim$ 5.5\% on Douban-book, and 1.1\% $\sim$ 8.2\% on ML-1M in terms of R@$K$ and N@$K$, where $K$ = 5, 10, and 20.
The results support our hypothesis that CaDRec is effective for (1) mitigating over-smoothing by incorporating sequential information into GCN operators, and (2) eliminating the impact of popularity while disentangling item representations from user individual biases. Table~\ref{tab:comparison} also promotes the following observations and insights:

\begin{itemize}[nosep,labelindent=0em,leftmargin=1em,font=\normalfont]

\item The OSA-based baselines attempt to address the over-smoothing issue by performing graph augmentations on nodes, edges, or structures. However, they are less effective than CaDRec for two possible reasons: (1) their GCN operators solely propagate messages on the connectivity between nodes, and (2) the representations learned by their graph learning are impeded by popularity and individual bias.

\item 
Both InvCF and DICE utilize two neural encoders to independentize the representations of preference and popularity for debiasing. In contrast, the CaDRec tokenizes historical item activeness as popularity representations via positional encoding. The LightGCL employs singular value decomposition (SVD) to align graph contrastive learning representations for popularity debiasing, disregarding the effective sequential context for GCNs.

\item Note that EEDN and HCCF explore beyond pair-wise relations between users and items via hypergraphs. However, EEDN, which additionally leverages sequential contexts, achieves remarkable performance, highlighting the effectiveness of sequential contexts in enriching user representations. In contrast, the results of SA-based models indicate that relying solely on sequential patterns is insufficient to provide accurate recommendations. 


\item 
The GCL-based methods, NCL and XSimGCL, are competitive among baselines. 
This implies that the contrast of pairs, e.g. a node and its neighbors, contributes to the generalizability of graph embeddings making nodes more distinctive.


\item 
The SGL, NCL, LightGCL, and XSimGCL introduce the InfoNCE loss, while they perform worse than the CaDRec, indicating advantages of the CaDRec loss: (1) it discriminatively encourages positive samples instead of penalizing stochastic candidates, which could be the false-negative items that users prefer, (2) it properly regularizes positive samples to avoid overfitting the popularity bias, and (3) it enables the injection of sequential contexts to enrich user representations.



\item 
Although DiffRec effectively denoises user\textendash item interactions via generative learning, it is fairly limited by insufficient prior knowledge, various contexts, and debiases.

\end{itemize}

\begin{table}
  \caption{\small{Ablation study. ``w/o X'' denotes the removed parts. ``SA'', ``ER'', ``WS'', and ``Dis'' indicate the SA correlation, the embedding regularization, the weighting scheme, and the disentanglement components, respectively. The best value of the hyperparameter of $\delta$ in Eq. (\ref{eq:hgc}) is 0 which is equal to excluding SA, leading to the same result. }}
  \label{tab:ablation}
\resizebox{1\columnwidth}{!}{ 
  \begin{tabular}{c|cc|cc|cc|cc} 
    \toprule
      \multirow {2}{*}{Model} &\multicolumn{2}{c|}{Yelp2018} &\multicolumn{2}{c|}{Foursquare} &\multicolumn{2}{c|}{ML-1M}&\multicolumn{2}{c}{Douban-book} \\
& R@20 & N@20 & R@20 & N@20 & R@20 & N@20 & R@20 & N@20 \\ 
    \midrule
    w/o SA & 0.0775 & 0.0649 & 0.1282 & 0.1039 & \textbf{0.2907} & \textbf{0.3261} & 0.1954 & 0.1846 \\
    w/o Dis & 0.0759 & 0.0632 & 0.1238& 0.1010 & 0.2593& 0.2913 & 0.1933& 0.1816  \\
    \midrule
    w/o ER & 0.0720 & 0.0605 & 0.1217 & 0.1002 & 0.2580 & 0.2921 & 0.1713 & 0.1632 \\
    w/o WS & 0.0751 & 0.0622 & 0.1273 & 0.1044 & 0.2795 & 0.3152 & 0.2038 & 0.1919 \\
    \midrule
    Full & \textbf{0.0792} & \textbf{0.0668} & \textbf{0.1314} & \textbf{0.1068} & \textbf{0.2907} & \textbf{0.3261} & \textbf{0.2085} & \textbf{0.1960} \\
  \bottomrule
\end{tabular}
}
\end{table}

\subsection{Ablation Study}

Table~\ref{tab:ablation} shows an ablation study to examine the effect of each module on the CaDRec, which provides the following observations:
\begin{itemize}[nosep,labelindent=0em,leftmargin=1em,font=\normalfont]
\item 
 We can see that in most cases, i.e., on the Yelp2018, Foursquare, and Douban-book datasets, the SA correlation as auxiliary information greatly enhances the ability to select effective neighbors and further alleviates the over-smoothing issue.

\item 
The comparative results of CaDRec with and without interaction disentanglement indicate that the disentanglement contributes to promoting performance. It poses the essentiality of decoupling items' embeddings from popularity influences and user individual bias for recommendations. 

\item 
The CaDRec without embedding regularization is affected by popularity bias, and its performance drops significantly, suggesting that regularization is effective for correcting gradient propagations to update embeddings. 

\item 
The results without a weighting scheme, i.e., ``w/o WS'', are in underperformance, indicating that balancing the gradients of updating IA and FIA item embeddings noticeably reinforces debiased representations for recommendations. 
\end{itemize}

\subsection{Further Analysis of CaDRec}

\begin{table}
  \caption{\small{Study of debiasing categories for disentanglement. ``Pop'', and ``Indi'' indicate popularity and individual biases, respectively.}}
  \label{tab:implicit}
\resizebox{1\columnwidth}{!}{ 
\begin{tabular}{cc|cc|cc|cc|cc}
\toprule
      \multicolumn{2}{c|}{Category} &\multicolumn{2}{c|}{Yelp2018} &\multicolumn{2}{c|}{Foursquare} &\multicolumn{2}{|c}{ML-1M} &\multicolumn{2}{|c}{Douban-book} \\
\midrule
      Pop & Indi & R@20 & N@20 & R@20 & N@20 & R@20 & N@20 & R@20 & N@20 \\   
\midrule
  \ding{56} &  \ding{56}& 0.0759 & 0.0632 & 0.1238& 0.1010 & 0.2593& 0.2913 & 0.1933& 0.1816  \\
  \ding{52} &  \ding{56}& 0.0771 & 0.0645 & \underline{0.1294}& \underline{0.1053} & 0.2764& 0.3119 & \underline{0.2023}& \underline{0.1904} \\
  \ding{56} & \ding{52} & \underline{0.0774} & \underline{0.0656} & 0.1258& 0.1027 & \underline{0.2783}& \underline{0.3137} & 0.2005& 0.1876 \\
  \ding{52} & \ding{52} & \textbf{0.0792} & \textbf{0.0668} & \textbf{0.1314}& \textbf{0.1068} & \textbf{0.2907}& \textbf{0.3261} & \textbf{0.2085}& \textbf{0.1960} \\
\midrule
\multicolumn{2}{c|}{\textbf{Improv.}}& \textbf{+4.7\%} & \textbf{+5.7\%} & \textbf{+6.1\%}& \textbf{+5.7\%} & \textbf{+12.1\%}& \textbf{+12.0\%} & \textbf{+7.9\%}& \textbf{+7.9\%}  \\
\bottomrule
\end{tabular}
}
\end{table}

\subsubsection{Effect of Disentanglement}
To verify the effectiveness of the disentanglement component, we obtained the experimental results on four datasets in Table~\ref{tab:implicit}. We can conclude that both the popularity and user-individual debiasing contribute to the performance of the CaDRec. It improved performance by 4.7\% $\sim$ 12.1\% in R@20 and 5.7\% $\sim$ 12.0\% in N@20 on average in four datasets. Specifically, lessening the popularity bias yields more benefits for the Foursquare and Douban-book datasets, while user-individual debiasing can better refine representations for the Yelp2018 and ML-1M datasets.

\subsubsection{Integration Effect of HGC and SA}

\begin{table}
  \caption{\small{Study of the integration of HGC and SA. ``WeightAdd''  denotes that HGC and SA are integrated by weighted addition.}}
  \label{tab:integration}
\resizebox{1\columnwidth}{!}{ 
  \begin{tabular}{c|cc|cc|cc|cc}      
\toprule
      \multirow {2}{*}{Method } &\multicolumn{2}{c|}{Yelp2018} &\multicolumn{2}{c|}{Foursquare} &\multicolumn{2}{c|}{ML-1M}&\multicolumn{2}{c}{Douban-book} \\
& R@20 & N@20 & R@20 & N@20 & R@20 & N@20 & R@20 & N@20 \\ 
\midrule
    Only-SA & 0.0766 & 0.0648 & 0.1195 & 0.0972 & 0.1692 & 0.1553 & 0.1921 & 0.1799 \\
    Only-HGC & \underline{0.0775} & \underline{0.0649} & \underline{0.1282} & \underline{0.1039} & \textbf{0.2907} & \textbf{0.3261} & \underline{0.1954} & \underline{0.1846} \\
 \midrule   
    WeightAdd & 0.0769 & 0.0648 & 0.1220 & 0.0986 & 0.2588 & 0.2946 & 0.1938 & 0.1826 \\
    Perturbation & \textbf{0.0792} & \textbf{0.0668} & \textbf{0.1314}& \textbf{0.1068} & \underline{0.2714} & \underline{0.3089} & \textbf{0.2085}& \textbf{0.1960} \\
\bottomrule
\end{tabular}
}
\end{table}

Table~\ref{tab:integration} shows how HGC and SA affect the overall performance. We can see that perturbation learning is effective for fusing structural and sequential correlations. Furthermore, we have the following observations: (1) the user\textendash item interactions are more suitable for modeling with structured graphs rather than sequential patterns, as HGC outperforms SA when used solely as an encoder, (2) comparing the results of ``only-HGC'' and ``perturbation,'' in most cases, injecting sequential information can enhance the convolution operator, (3) a simple weighted addition of the features from two different feature spaces always makes the performance worse, indicating that considering their distributions is essential during feature fusion, and (4) interestingly, the empirical results on the ML-1M dataset demonstrate that 
if the standalone SA exhibits inferior performance on a given dataset, it is not appropriate to integrate the SA in this dataset. 


\begin{figure}
  \includegraphics[width=\linewidth]{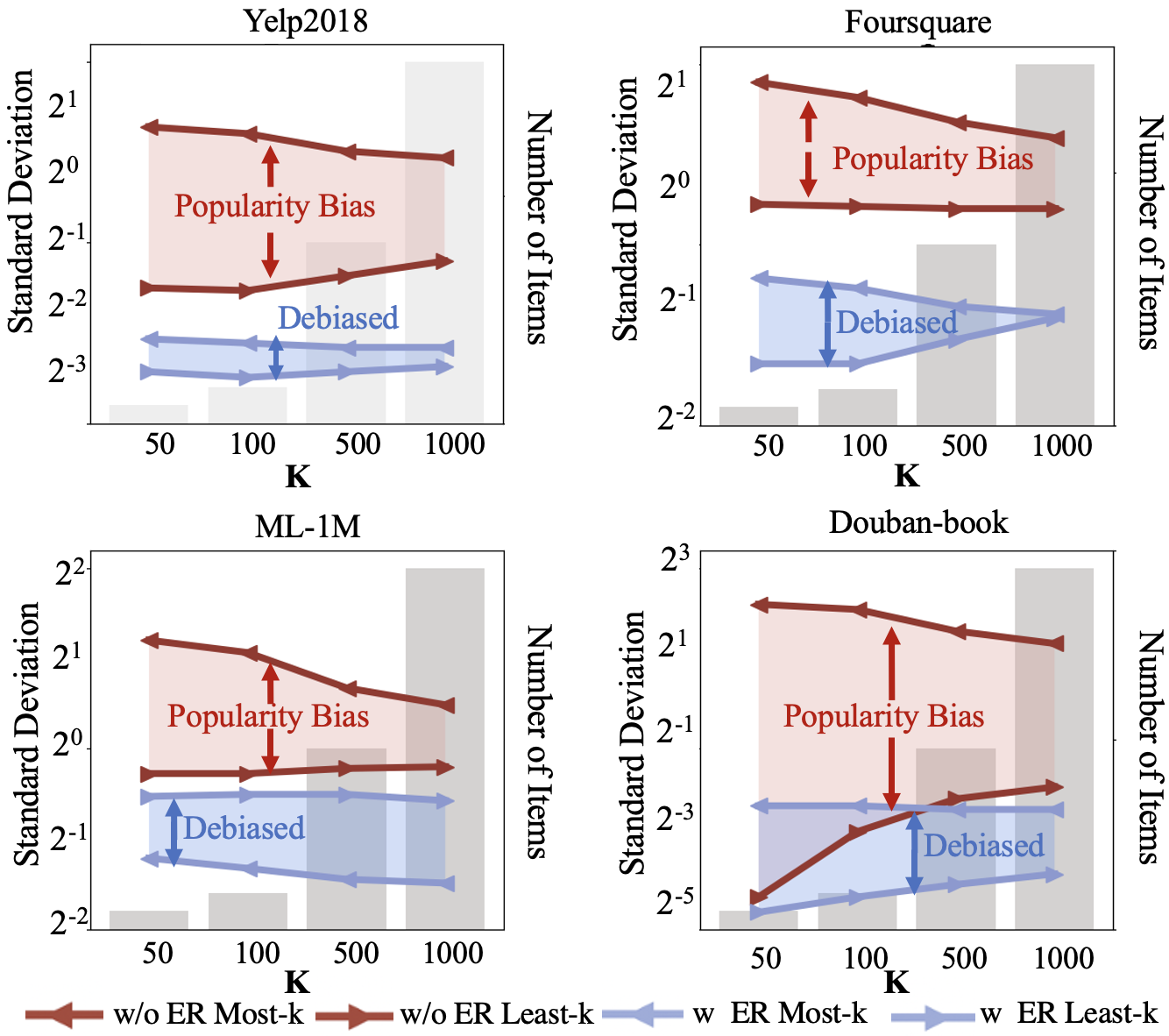}
  \caption{{Effect of embedding regularization. The Y-axis represents exponential growth with a base of 2.}}
  \label{fig:popbias_st}
\end{figure}

\subsubsection{Effect of Debiasing on Embeddings}
We further investigated the standard deviations (SD) of the item embeddings to evaluate the effectiveness of embedding debiasing. More specifically, we chose the most and least popular $k$ items, where $k$ $\in$ \{50, 100, 500, 1000\}, and examined the SD of their embeddings with and without regularization, respectively. As illustrated in Fig.~\ref{fig:popbias_st}, the result without regularization would show a large gap between embeddings of popular and unpopular items, resulting in the model tending to recommend the most popular items over users' pure preferences. 
It is noteworthy that regularization significantly narrows this gap and achieves notable improvement for the four datasets in Table. \ref{tab:ablation}, indicating its effectiveness in alleviating popularity bias.

\subsubsection{Effect of Weighting Scheme}
\begin{figure}
  \includegraphics[width=\linewidth]{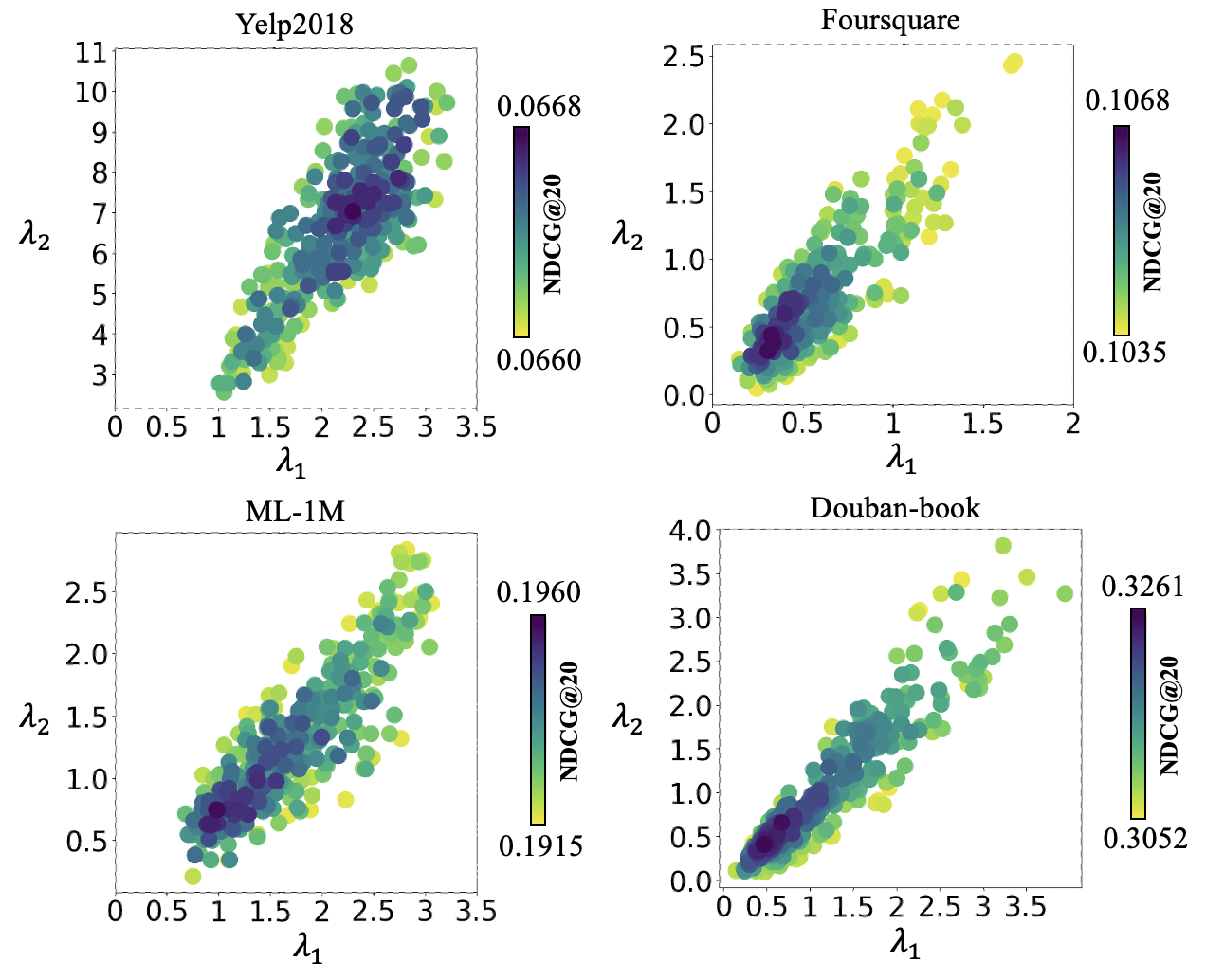}
  \caption{Effect of various weighting schemes.}
  \label{fig:beta_lamdba}
\end{figure}

Fig.~\ref{fig:beta_lamdba} illustrates how the weighting scheme affects the CaDRec. 
We can see that various weighting schemes to balance embedding gradients of IA and FIA items significantly influence the quality of recommendations.
We can also observe that (1) it suggests an optimal linear dependency between the optimal $\lambda_1$ and $\lambda_2$, making it easier to seek the optimal solution when deploying a new dataset, and
(2) the scaling coefficient affects the performance because even if values of $\lambda_1$ and $\lambda_2$ are close to the optimal linear region, their scaling still affects the recommendation quality of four datasets.

\noindent

\begin{figure}[t]
  \includegraphics[width=\linewidth]{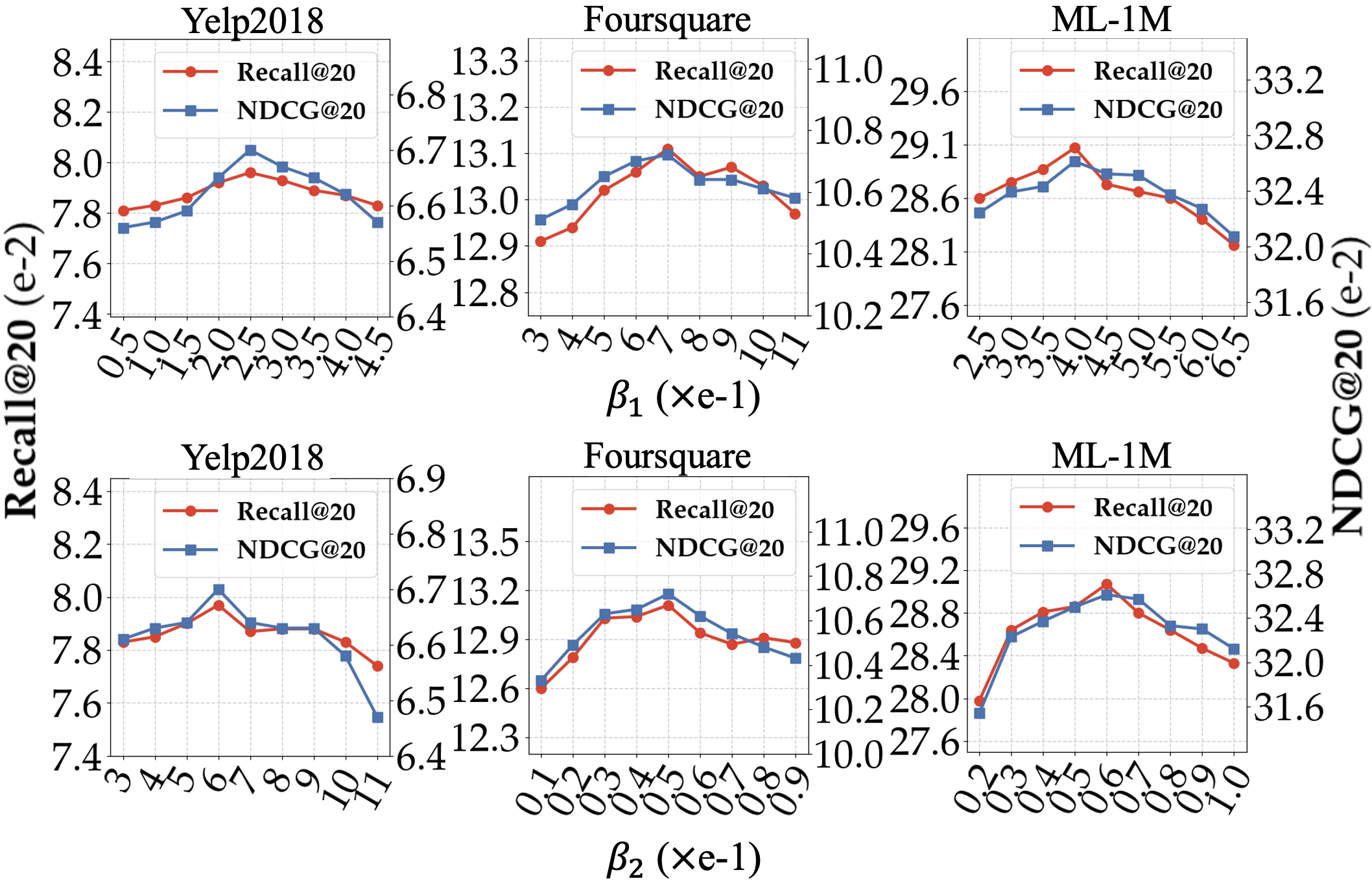}
  \caption{Effect of hyperparameters $\beta_1$ and $\beta_2$.}
  \label{fig:beta}
\end{figure}

\subsubsection{Study of Hyperparameter Sensitivity}
In this part, we investigate the impact of two key hyperparameters, i.e., $\beta_1$ and $\beta_2$, in CaDRec. Due to the limited
space, we only report the results on the Yelp2018, Foursquare, and ML-1M datasets, and the observations on the Douban-book dataset are similar.

\noindent
\textbf{Effect of $\beta_1$.} 
The hyperparameter of $\beta_1$ was assigned to vary the influence level of popularity. We can see from Fig.~\ref{fig:beta} that the best values of $\beta_1$ are 0.25 for Yelp2018, 0.7 for Foursquare, and 0.4 for ML-1M, respectively. Overall, the sensitivity of $\beta_1$ depends on the datasets, as there is no significant fluctuation by R@20 and N@20 on Yelp2018 and Foursquare datasets compared to the ML-1M dataset.

\noindent
\textbf{Effect of $\beta_2$.} 
The hyperparameter of $\beta_2$ is to control the penalization level of the L$_2$ regularization. Fig.~\ref{fig:beta} shows that the best values of $\beta_2$ are 0.6 for Yelp2018, 0.05 for Foursquare, and 0.06 for ML-1M, respectively.
We can conclude that a large average number of user interactions, for example, 68.2 in Foursquare and 95.3 in ML-1M, requires a small value of $\beta_2$ because more item embeddings are penalized in each iteration.



\begin{figure*}[t]
  \includegraphics[width=\linewidth]{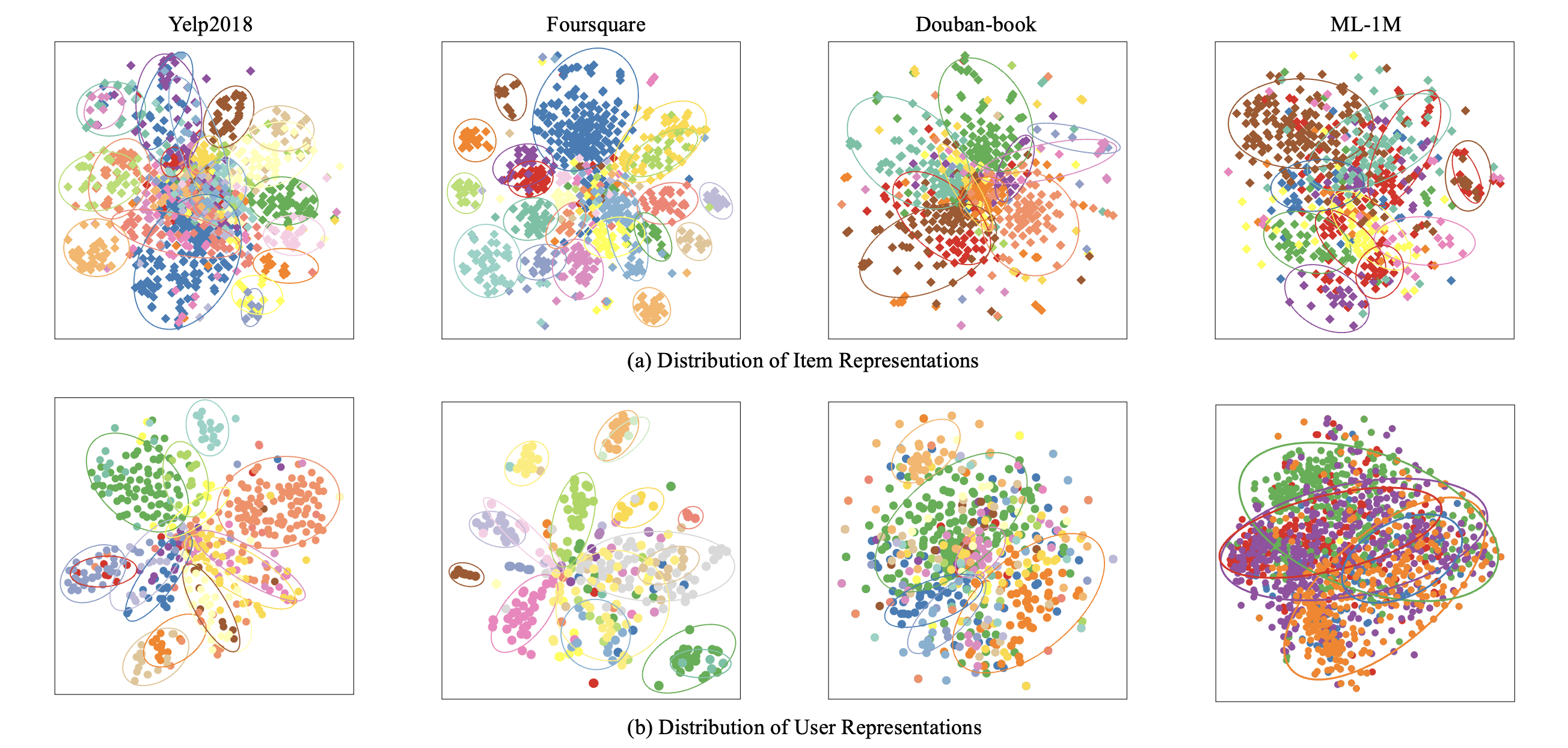}
  \caption{Visualization of user and item representations. Each color denotes a user interacting with the same item or items that interact with the same user.}
  \label{fig:visualization}
\end{figure*}  
\subsubsection{Visualization of User and Item Representations}


To verify the effectiveness of the CaDRec, we visualized distributions of user and item representations learned by the CaDRec using t-SNE~\cite{van2008visualizing}, as illustrated in Fig.~\ref{fig:visualization}. We have the following observations:

\begin{itemize}[nosep,labelindent=0em,leftmargin=1em,font=\normalfont]

\item Overall, the dots in the same color are close to each other, indicating that the CaDRec effectively differentiates similar users and items through learning contextualized and debiased representations from user\textendash item interactions.



\item The visualization effect depends on datasets, i.e., the strength relationship is Yelp2018 $\approx$ Foursquare $\textgreater$ Douban-book $\textgreater$ ML-1M. One possible reason for this is, as shown in Table. \ref{tab:dataset}, that user and item representations could be more specific when the number of candidates increases, as reflected in Fig.~\ref{fig:visualization}.

\item Surprisingly, the visualization result on the ML-1M dataset is worse than the others, while Table~\ref{tab:comparison} shows that the CaDRec works well on this dataset. The reason is that most users have interacted with many of the same items in the dataset according to the average interaction count in Table \ref{tab:dataset}, leading to their representations sticking together in the 2D semantic space.


\end{itemize}

\subsubsection{Time Complexity}
We selected the top three baselines to compare the time complexity which highlights the efﬁciency of the CaDRec against the three baselines. We have the following observations and insights: (1) XSimGCL and NCL regard users and items as equivalent nodes in GCN which result in the complexity of $\mathcal{O}((M+N)^2)$, while that of HGC by the CaDRec is $\mathcal{O}(L^2)$, where $(M+N) \gg L$; (2) the BPR loss ($\mathcal{O}(MLd_{neg})$) of XSimGCL and NCL uses $d_{neg}$ negative samples to train each interaction of users, while the CaDRec learns all interactions of a user at a time; (3) XSimGCL and NCL regard the other samples in the same batch as negative samples for contrastive learning ($\mathcal{O}(MLB)$); and (4) EEDN utilizes a stack of Transformers ($\mathcal{O}(L^2)$) in its decoder to capture implicit features. This leads to an additional exponential increase in time complexity for $L$ compared to CaDRec.

\begin{table}
\centering
  \caption{{Comparison of time complexity in various components. $B$: batch size. ``Obj. Loss'' and ``CL Loss'' refer to the objective loss and contrastive learning loss, respectively.}}
  \label{tab:complexity}
\resizebox{\columnwidth}{!}{ 
  \begin{tabular}{c|c|c|c|c}      
\toprule
      Component & XSimGCL & NCL& EEDN & \textbf{CaDRec} \\
\midrule
    Encoder & $\mathcal{O}((M+N)^2)$ & $O((M+N)^2)$ & \ \ \ \ \  \ \ $\mathcal{O}(L^2)$ \ \  \ \ \ \ & \ \ \ \ \ \ $\mathcal{O}(L^2)$ \ \ \ \ \ \ \\
    Decoder & $\mathcal{O}(N)$ & $\mathcal{O}(N)$ & $\mathcal{O}(L^2+N)$ & $\mathcal{O}(N)$ \\
    Obj. Loss & $\mathcal{O}(MLd_{neg})$ & $\mathcal{O}(MLd_{neg})$ & $\mathcal{O}(M)$ & $\mathcal{O}(M)$ \\
    CL Loss & $\mathcal{O}(MLB)$ & $\mathcal{O}(MLB)$ & - & - \\
\bottomrule
\end{tabular}
}
\end{table}

%% file: contents/Conclusion.tex
We proposed a CaDRec for a recommendation that integrates HGCs and Transformers to mitigate the over-smoothing issue and disentangle user\textendash item interactions for debiasing. We also showed that the imbalance of the gradients for updating item embeddings magnifies the popularity bias and provides solutions. Extensive experiments and analyses have demonstrated the efficacy and efficiency of our CaDRec. In the future, we will (i) explore more effective ways to balance the gradients for updating item embeddings, and (ii) extend the CaDRec to automatically exclude ineffective contexts such as sequential contexts on the ML-1M dataset.

%% file: contents/Acknowledgements.tex
We would like to thank anonymous reviewers for their thorough comments. This work is supported by NEC C\&C (No.24-004), China Scholarship Council (No.202208330093), and JKA (No.2023M-401).